# Electric Field Tuning of the Surface Band Structure of Topological Insulator $Sb_2Te_3$ Thin Films

Tong Zhang[1,2], Jeonghoon Ha[1,2,3], Niv Levy[1,2], Young Kuk[3], Joseph Stroscio[1*]

[1]Center for Nanoscale Science and Technology, NIST, Gaithersburg, MD 20899, USA
[2]Maryland NanoCenter, University of Maryland, College Park, MD 20742, USA
[3]Department of Physics and Astronomy, Seoul National University, Seoul 151-747, Korea

We measured the response of the surface state spectrum of epitaxial $Sb_2Te_3$ thin films to applied gate electric fields by low temperature scanning tunneling microscopy. The gate dependent shift of the Fermi level and the screening effect from bulk carriers vary as a function of film thickness. We observed a gap opening at the Dirac point for films thinner than four quintuple layers, due to the coupling of the top and bottom surfaces. Moreover, the top surface state band gap of the three quintuple layer films was found to be tunable by back gate, indicating the possibility of observing a topological phase transition in this system. Our results are well explained by an effective model of 3D topological insulator thin films with structure inversion asymmetry, indicating that three quintuple layer $Sb_2Te_3$ films are topologically nontrivial and belong to the quantum spin Hall insulator class.

Topological insulators (TI) represent a novel state of quantum matter that has topologically protected edge or surface states [1,2]. Theory and experimental verification of TI matter rapidly increased since the first predictions of these states of matter in 2-dimensions (2D). 2D TIs are equivalent to quantum spin Hall (QSH) insulators which host 1-dimensional (1D) spin polarized edge states [3,4]. The concept of QSH insulators has been generalized to three-dimensional (3D) TIs, which are 3D band insulators surrounded by 2D metallic surface states with helical spin texture [1,2]. These surface states are protected by time reversal symmetry and are not eliminated by scattering from weak non-magnetic disorder. Recently, the family of bismuth chalcogenide materials ($Bi_2Se_3$, $Bi_2Te_3$ and $Sb_2Te_3$) have been confirmed to be 3D TIs with a single Dirac cone at the $\Gamma$ point [5–11]. In sufficiently thin 3D TI films the top and bottom surface states hybridize, disrupting the Dirac cone and creating an energy gap at $k=0$, allowing a range of topologically interesting phases [12–21]. In the ultrathin film geometry, the low-lying physics is described by two degenerate Dirac hyperbolas. Each massive band has a $k$-dependent spin configuration determined by the energy gap and spin-orbit coupling. As a result, ultrathin 3D TI films can be considered a 2D QSH system with 1D spin helical edge states if the system is topologically nontrivial. Calculations show that the surface band gap of the ultrathin 3D TIs displays an oscillating dependence on thickness [13–21], and the system may alternate between being a topologically trivial or nontrivial insulator. More interestingly, the band topology is predicted to be tunable by external electric fields, which cause structure inversion asymmetry (SIA), leading to the possibility of controlling a phase transition between topologically trivial and nontrivial phases [15–21]. For practical applications it is important to study the surface state properties of different phases. This may enable control of the QSH edge states with potential applications in future spintronic devices.

In this letter, we study the gate tunable electronic structure of the 3D TI $Sb_2Te_3$ using low temperature scanning tunneling microscopy and spectroscopy (STM/STS) on *in-situ* epitaxially grown thin films on back gated $SrTiO_3$ substrates. We find that the field effect gating of the thin films varies strongly with film thickness. This was determined by measuring the shift of the thin film quantum well states with back gate potential. For films below four quintuple layers (QL) thickness a band gap is observed in the surface state spectrum, which we attribute to the hybridization of the top and bottom surfaces. At a film thickness of 3 QL the field effect is able to reduce the gap with increased gate field. Comparing an effective Hamiltonian model with structure inversion asymmetry to the gap dependence on gating implies that 3 QL $Sb_2Te_3$ film is topologically non-trivial, and suggests that a topological phase transition may occur if the electric field applied to the film is large enough.

Despite many transport measurements on gated TI devices [22–26], making a gate tunable TI film that is accessible to low temperature STM is challenging, mainly because chalcogenide TI's surfaces are environmentally sensitive and degrade easily through exposure to air during standard lithography processes. We overcome this problem by epitaxially growing TI films on pre-patterned $SrTiO_3$ (111) (STO) substrates mounted on special sample holders [27]. The samples were transferred in ultra-high vacuum into an STM right after growth, which avoided any *ex-situ* processing. The sample bias is applied through two top electrodes and a gate voltage is applied to a back contact on the STO. STM/STS and two-terminal transport measurements are both carried out *in-situ*, in a homemade STM operating at 5 K, which is connected to the molecular beam epitaxy (MBE) system [28]. The samples studied here are $Sb_2Te_3$ thin films grown by co-depositing Sb and Te at a substrate temperature of 200 °C. Samples were characterized with reflection-high-energy-electron diffraction during growth.

Fig. 1(a) shows the typical topography of a nominally 5 QL thick $Sb_2Te_3$ film. Terraces with thicknesses between 2 QL to 5 QL can be found on the same film, and their height determined by finding pinholes to the STO substrate level. Figure 1(b) shows atomic resolution of the Te terminated (111) surface with lattice constant of 0.42 nm. The differential tunneling conductance, *dI/dV*, is measured by standard lock-in techniques, with a modulation frequency of ≈500 Hz. Typical *dI/dV* spectra on different layer thicknesses are shown in Fig. 1(c). The pronounced peaks (see arrows in Fig. 1(c)) are attributed to quantum well states (QWS) from bulk bands that undergo quantum confinement in a thin film geometry. [29]. A closer inspection of the surface state region, obtained at lower junction impedance, shows a hybridized gap at the sample bias ≈$V_B$=0.25 V on 2 QL and 3 QL terraces [Fig. 1(d)], with a width of ≈160 meV and 80 meV, respectively. The gap is absent for the 4 QL film and leaves a "V" shaped density of states that indicates an intact Dirac cone. Similar results were reported on $Sb_2Te_3$ film grown on graphitized SiC [30].

Our experimental setup allows, for the first time, for a gate voltage $V_G$ to be applied to the $Sb_2Te_3$ film during STM measurements, without any *ex-situ* processing and exposure of the film to atmosphere. To measure the gating characteristics, we charged the sample (which can be thought of as a TI-STO-gate capacitor) with a current source while measuring the change of $V_G$. Fig. 1(e) shows the total charge density *n* and displacement field versus $V_G$ by integrating the current over time. *n* reaches about $4.5 \times 10^{13}/cm^2$ at $V_G$ = 200 V, which indicates a large gating ability of STO at low temperature. A two-terminal film resistance measurement versus $V_G$ is shown in Fig. 1(f), which displays a rapid increase at positive $V_G$, as expected for gating a *p*-doped semiconductor.

The gating effect on the local density of states (LDOS) is measured using the thin film QWS as a fiduciary mark in the tunneling spectra [Fig. 1(c)]. The overall gating dependence of the LDOS is shown in Figs. 2(a) and 2(b) for 2 QL and 3 QL films by focusing on a large sample bias range (-0.4 V to 0.4 V). Increasing gate voltage is accompanied by the whole spectrum moving to lower energies, which is a clear signature of the shifting of the Fermi energy $E_F$ due to the gate induced doping of the $Sb_2Te_3$ film. Note in the tunneling experiment $E_F$ is at $V_B = 0$ V, and therefore the change in doping is observed by the shift of the electronic bands relative to zero sample bias. The relative shifts of $E_F$ (with respect to $V_G=0$) as a function of displacement field are plotted in Fig. 2(c) for film thicknesses of 2 QL, 3 QL, and 4 QLs. One can see that the gating tunability decreases fast with increasing film thickness, approximately as $d^{-2}$ [Fig. 2(d)]. For the 2 QL film, the thinnest case, we estimate the change in surface carrier density, corresponding to the measured shift in $E_F$, only reaches to ¼ of the total charge density induced by the gate. Therefore the majority of the carriers are expected to be in the bulk of the film, which screens the electric field reaching the top surface. Subsequently, in these thin films an overall Fermi level shift and band bending will coexist through the film providing a potential asymmetry between the top and bottom surfaces [31]. Nevertheless, the gate tunability is still sufficient to deduce the topological character of the 3 QL film, as described below.

Now we focus on the hybridization gaps that open due to the coupling of the top and bottom surfaces, and their response to applied fields. Figures 3(a) and 3(b) show the tunneling spectra of the surface state band gaps for 2 QL and 3 QL terraces, respectively, as a function of $V_G$. The gap size is measured by the peak positions in the second derivative $d^2I/dV^2$, which indicate the inflection points on either side of the surface state gap [see dashed line in Fig. 3(a)].

A noticeable feature is that the surface states gap is rather constant at 2 QL, but varies considerably with $V_G$ for 3 QL. Differences in the response of the gap to fields, as observed between the 2 QL and 3 QL films, can occur due changes in the topology of the system. Such changes can arise with variations as small as 1 QL in film thickness [18,19]. The gap for the 3 QL film shows a linear dependence on displacement field [Fig. 3(d)]. This is reminiscent of bi-layer graphene where the asymmetric potential controls the gap at the Dirac point [32,33]. The difference here is that in a TI thin film, the inter-layer and spin orbit coupling play important roles.

To understand how an electric field affects the surface band structure, we refer to an effective Hamiltonian model that describes thin 3D TI films as [17,18],

$$H_{eff}(k) = E_0 - Dk^2 + \hbar v_F \left(k_y \hat{\sigma}_x - k_x \hat{\sigma}_y\right) + \left(\frac{\Delta}{2} - Bk^2\right)\hat{\tau}_z \hat{\sigma}_z + \gamma U \hat{\tau}_x \qquad (1)$$

The first three terms in Eq. (1) account for isolated TI surface states. The fourth term describes the inter-surface coupling, which induces a hybridized gap $\Delta$ and parabolic term $(Bk^2)$. $\hat{\sigma}$ and $\hat{\tau}$ are Pauli matrices of electron spin and isospin (*i.e.* bonding/anti-bonding states of the top and bottom surfaces), respectively. The asymmetric potential leading to structure inversion asymmetry (SIA) is introduced by adding an effective potential energy $\gamma U$ to $H_{eff}$. The parameters $E_0, D, B, \Delta, v_F$ and $\gamma$ are material and thickness dependent, and $U$ is the potential difference between the top and bottom surfaces induced by the gate electric field. Solving for the energy eigenvalues for the effective Hamiltonian in Eq. (1) gives rise to four surface state energy bands [17,18],

$$E_{1\pm} = E_0 - Dk^2 \pm \sqrt{\left(\frac{\Delta}{2} - Bk^2\right)^2 + \left(|\gamma U| + \hbar v_F k\right)^2} \qquad (2)$$

$$E_{2\pm} = E_0 - Dk^2 \pm \sqrt{\left(\frac{\Delta}{2} - Bk^2\right)^2 + \left(|\gamma U| - \hbar v_F k\right)^2} \qquad (3)$$

where the $+(-)$ sign stands for the conduction (valence) band, and the 1 (2) stands for the inner (outer) branches of the bands. Without the SIA term ($\gamma U = 0$) these states consist of spin degenerate conduction and valence bands separated by a hybridization gap $\Delta$ [Fig. 4(a)]. The presence of applied potential $U$ leads to a Rashba-like splitting of conduction and valence bands [Figs. 4(b)]. Moreover, from Eqs. (2) and (3), the actual gap size also varies with potential asymmetry [see Figs. 4, (a)-(d)]. This is directly related to our observations of the 3 QL $Sb_2Te_3$ gap decreasing with increasing applied gate potential [Figs. 3, (b) and (d)], which is examined below.

As discussed in Refs. [17] and [18], the sign of $\Delta/B$ and the value of the SIA term determine if the system is a topologically trivial or non-trivial insulator. If $B^2 > D^2$ and $\Delta$ and $B$ have the same sign, *i.e.* $\Delta/B > 0$, the system is in the quantum spin hall state. The SIA term $\gamma U$ reduces the energy gap [Fig. 4(b)] and eventually close it at the critical potential difference $\gamma U_C = \hbar v_F \sqrt{\Delta/2B}$ [Fig. 4(c)] and then reopen it with increasing potential [Fig. 4(d)], leading to a topological phase transition as a function of applied electric field [17,19]. If $\Delta$ and $B$ have opposite sign, the system remains gapped as a function of applied electric field. Fig. 4(f) compares these two cases, which we use below to determine the topological character of the 3 QL film. We note that $\gamma U$ contributes to Eq. (2) and (3) through its absolute value, so the gap variation should be symmetric with respect to the gate voltage. However, in Fig. 3(d) the gap displays a monotonic dependence on displacement field within the range investigated. We expect this lack of symmetry is due to an initial band bending which exists even at $V_G = 0$ [see schematic in Fig. 4(e)] and is common for hetero-junctions.

We can estimate the initial bending direction from the thickness dependence of the position of the Dirac point (or mid-gap energy position). As shown by the dashed line in Fig. 1(d), the middle of the gap moves closer to $E_F$ with increasing thickness from 2 QL to 4 QL, which implies the top surface of the film is less doped at increased thicknesses. The increased doping for thinner films is probably due to an increased number of defects at the TI/STO interface. These defects are very effective acceptors [29]. The higher hole doping at the interface gives an upward band bending through the film. The band bending is present at zero applied gate potential [Fig. 4(e)] and will affect the surface band gap as if a gate field where applied [31]. Applying a negative gate potential leads to hole accumulation and further enhances the band bending intensity, and hence the potential asymmetry [Fig. 4(e) top panel]. We expect that the potential difference between the top and bottom 3 QL film is no less than the difference of the $E_F$ shifts of the top surfaces of the 2 QL and 3 QL films, *i.e.* 45 meV at $V_G =$ 200 V. Therefore, if the system is non-trivial, applying a negative gate potential leads to a decreased gap according to the above analysis [Fig. 4(f)]. This is what we observed in Fig 3(b) for 3 QL. We model the spectra in Fig. 3(b) for 3 QL by calculating the density of states from Eqs. (2) and (3), taking $v_F = 2.5$ eV Å, $B = 15$ eV Å$^2$, and $D = -5$ eV Å$^2$ from fits to photoemission spectroscopy and density functional calculation results in Ref. [34]. The model DOS curves are convoluted with a Gaussian (with standard deviation σ=20 meV) to account for instrumental broadening and are shown in Fig. 3(c). The fitting parameters used in the DOS calculation are: $\Delta = 0.16$ eV, with $E_0$ varying from 0.24 eV to 0.27 eV, and $\gamma U$ varying from 0.07 eV to 0.12 eV, as $V_G$ changes from 200 V to -200 V. The variation of $\gamma U$ is within the range of total potential asymmetry we achieved by gating ($\approx 50$ meV). The simulated spectra in Fig. 3(c) are in good agreement with experiment [Fig. 3(b)]. The lack of an observed maximum in the experimental gap measurements indicates that the initial band bending at $V_G = 0$ is so strong that even $V_G = 200$ V cannot flatten it. From the model calculation in Fig. 3(c), we observe the gap decreasing with increased potential asymmetry in agreement with the experimental observations. As shown in Fig. 3(d), the band gaps determined from the calculation show a trend quantitatively similar to the experiment for the range of potential asymmetry chosen for the model calculation. We note that the experimental spectra have a different background than the model DOS, which is likely from contributions from bulk states.

From the model parameters, $\Delta$ and $B$ have the same sign, $\Delta/B > 0$, which indicates a topologically nontrivial QSH phase. We determine that the 3 QL $Sb_2Te_3$ system is in a nontrivial topological state by noting that the band gap decreases with negative applied gate voltage. For the initial p-type doping, such a decrease can only occur if $\Delta$ and $B$ have the same sign [see Fig. 4(f)]. The topological character of the system is typically characterized by the $Z_2$ invariant, where $Z_2=1$ signifies the topological nontrivial phase [1,2]. Figure 4(g) summarizes the phase diagram in terms of the expected $Z_2$ invariant deduced from our results; at $V_G=0$, 3 QL $Sb_2Te_3$ is in a topologically nontrivial phase and thus should undergo a phase transition with increasing gate field. This conclusion is similar to what has been predicted for 4 QL $Sb_2Te_3$ thin films [19]. Gapless edge states are supposed to exist in this regime [13,17,19] and searching for these edge states will be an interesting subject for further studies. We expect that the topological phase transition should also be observable in future studies with more insulating material to obtain higher electric fields.

**Acknowledgements**: We thank Mark Stiles, Paul Haney, Minsung Kim and Jisoon Ihm for useful discussions. We thank Steve Blankenship and Alan Band for technical assistance. N. L., T. Z., and J. H. acknowledge support under the Cooperative Research Agreement between the University of Maryland and the National Institute of Standards and Technology Center for Nanoscale Science and Technology, Grant No. 70NANB10H193, through the University of Maryland. J. H. and Y. K. are partly supported by Korea Research Foundation through Grant No. KRF-2010-00349.

**Figure Captions**

Figure 1: Characterization of $Sb_2Te_3$ thin films. (a) STM topographic image, 150 nm × 150 nm, of nominally five-QL $Sb_2Te_3$ film grown by MBE on $SrTiO_3(111)$. (b) Atomic resolution STM image, 10 nm × 10 nm of $Sb_2Te_3$ film. The STM topographic height is shown in a color scale from dark to bright covering a range of 5 nm for (a) and 0.14 nm for (b). Tunneling parameters: $I$=30 pA, $V_B$=1.5 V for (a) and $V_B$ =0.2 V for (b). (c)-(d) $dI/dV$ spectra as a function of $Sb_2Te_3$ film thickness for $V_G$=0 V, for wide (c) and narrow (d) energy ranges. The narrow energy range spectra are obtained with lower tunneling impedance to boost signal to noise in the band gap region. The curves are offset vertically for clarity. The dashed line in (d) indicates the shift of the mid-gap position toward the Fermi level with increased thickness. (e) Total charge density and displacement field versus gate voltage for a 3 QL $Sb_2Te_3/SrTiO_3$ device obtained by measuring the induced voltage as the device was charged at constant current. (f) 3 QL $Sb_2Te_3$ film two terminal resistance versus gate voltage.

Figure 2: *In-situ* measurement of electric field gating of MBE grown $Sb_2Te_3$ thin films. $dI/dV$ versus $V_B$ spectra as a function of the indicated back gate voltage for a (a) 2 QL thick film, and (b) 3 QL thick film. A shift of the QW peaks to higher energy is seen with increased gate potential, as indicated by the dashed lines which are guides to the eye. (c) The relative shift of the QW peak positions versus displacement field as the gate potential is varied for 2, 3, and 4 QL films. Error bars are one standard deviation uncertainties in determining the QW peak positions. The solid lines are linear fits to the data points. (d) The gating tunability versus film thickness obtained from the slopes of the linear fits of the QW peak shifts in (c). The error bars are one standard deviation uncertainties from the linear fits. The solid line is a polynomial fit to the data points.

Figure 3: Electric field effect of surface state band gaps in MBE grown $Sb_2Te_3$ thin films. $dI/dV$ versus $V_B$ spectra of the surface state gap as a function of the indicated back gate voltage for (a) 2 QL $Sb_2Te_3$ film, and (b) 3 QL $Sb_2Te_3$ film. The curves are offset for clarity. (c) Calculated surface state DOS using the effective Hamiltonian in Eq. (1) and resulting dispersions in Eqs. (2) and (3) as a function of applied potential. See text for model parameters. (d) Measured surface state band gap from the spectra in (b) of 3 QL $Sb_2Te_3$ thin films as a function of displacement field (blue round symbols), and surface state band gap from the model DOS calculation in (c) versus $\gamma U$ (red square symbols). The band gaps are determined from the peak minimum and maximum in the 2nd derivative spectra (see green dashed curve in (a)) for both the experiment and model calculation. The error bars are one standard deviation uncertainty determined from the uncertainty in fitting the peaks. The solid line is a linear fit to the data with a slope of 17.3 ± 2.5 meV/V Å$^{-1}$. The error is one standard deviation uncertainty determined from the linear fit.

Figure 4: Topological phase transitions in applied electric field. (a)-(d) Surface state energy versus momentum dispersion calculated using Eqs. (2) and (3) showing the variation of the band gap as a function of increasing structure inversion asymmetry potential. See text for model parameters. At the potential $\gamma U$=0.18 =$\gamma U_C$ the gap closes indicating a topological phase

transition. (e) Schematic of band bending through the TI film. The diagram shows the conduction and valence bands going through the film. At each interface the surface state hyperbolic dispersions are indicated. At $V_G=0$ there is initial band bending indicating an initial potential asymmetry $\gamma U_0$ which increases with negative gate voltage. (f) Surface state band gaps calculated from the 3D TI model using Eqs. (2) and (3) with $\Delta=0.16$ eV, $D=-5$ eV Å$^2$, $v_F=2.5$ eV Å, and $\Delta/B>0$ (blue; $B=15$ eV Å$^2$), and $\Delta/B<0$ (red; $B=-15$ eV Å$^2$). (g) Schematic of the $Z_2$ topological invariant for 3 QL $Sb_2Te_3$ thin films as deduced from the analysis in this paper. The schematic shows the initial potential asymmetry position, and the topological phase transition that is approached with increasing negative gate voltage leading to a decrease of the surface state band gap as seen in (a) to (c) and in the experimental data in Figs. 3, (b) and (d). The grey bar indicates the approximate range of potential asymmetry probed by the current experiment [see Fig. 3(d)].

# References

* Author to whom correspondence should be addressed. joseph.stroscio@nist.gov

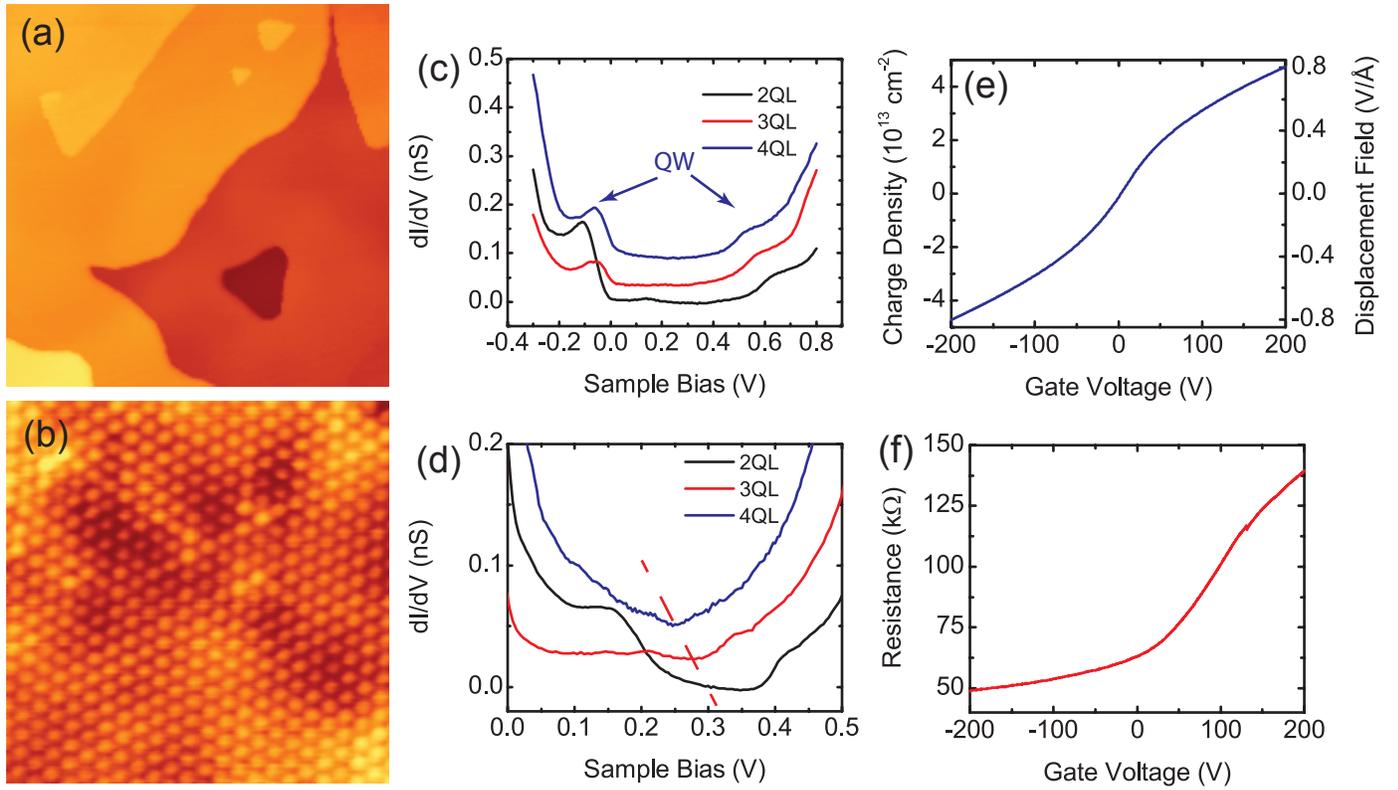

**Figure 1**. Zhang et al.

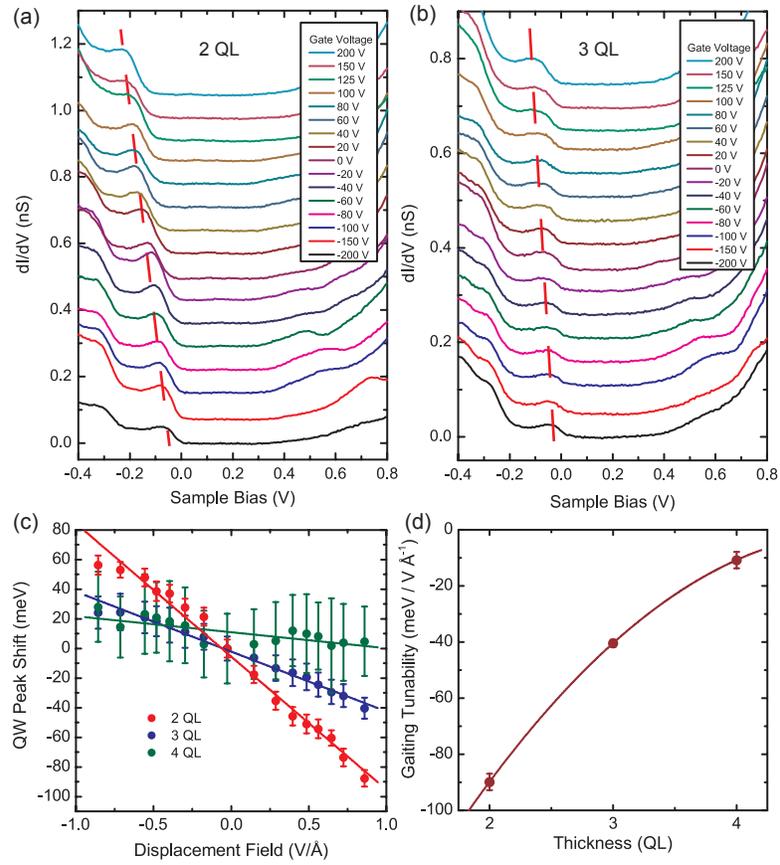

**Figure 2:** Zhang et al.

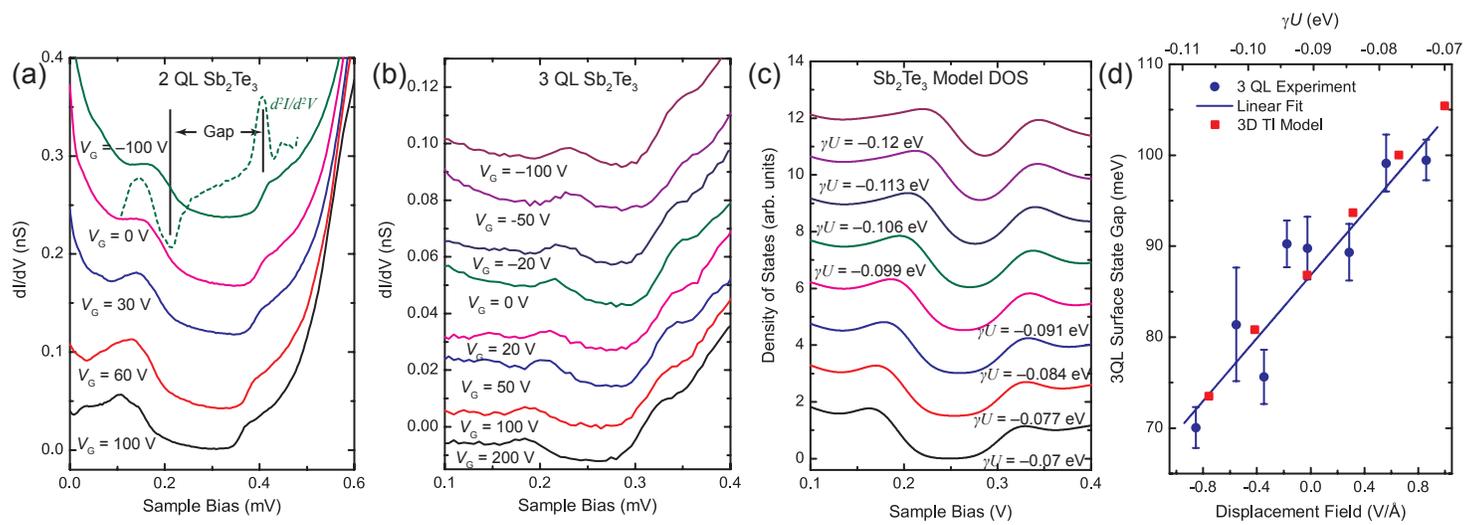

**Figure 3**. Zhang et al.

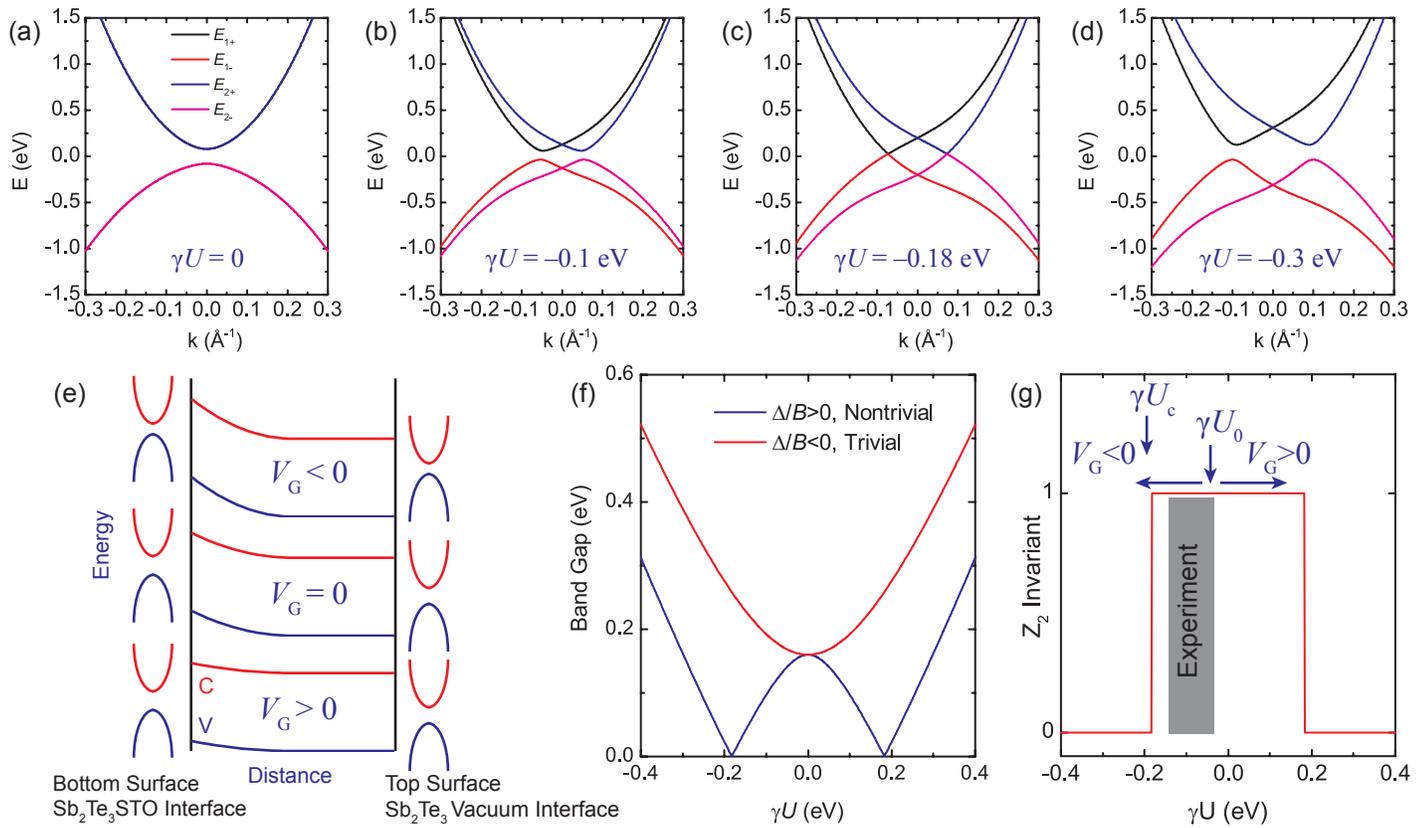

**Figure 4**. Zhang et al.